\documentclass[preprint,aps,nofootinbib,showpacs,amsfonts,epsf]{revtex4-2}

\input{epsf.tex}

\newcommand{\E}{{\cal{E}}}
\newcommand{\s}{\sigma}

\renewcommand{\a}{\alpha}
\renewcommand{\k}{\kappa}

\newcommand{\be}{\begin{equation}}
\newcommand{\ee}{\end{equation}}
\newcommand{\bea}{\begin{eqnarray}}
\newcommand{\eea}{\end{eqnarray}}
\newcommand{\ba}{\begin{array}}
\newcommand{\ea}{\end{array}}
\def\J#1#2#3#4{{#1} {\bf #2}, #3 (#4)}
\def\PRD{Phys. Rev. D}
\def\PR{Phys. Rev.}
\def\PRL{Phys. Rev. Lett.}
\def\PTP{Prog. Theor. Phys.}
\def\LRR{Living Rev. Relativ.}

\def\AJ{Astrophys. J.}

\def\MNRAS{Mon. Not. R. Astron. Soc.}
\def\JMP{J. Math. Phys. (N.Y.)}

\def\CQG{Class. Quantum Grav.}

\def\GRG{Gen. Relativ. Grav.}
\def\PLA{Phys. Lett. A}
\def\PLB{Phys. Lett. B}

\begin{document}
\draft
\title{Exterior field of neutron stars: The singularity structure\\ of vacuum and electrovac solutions}

\author{V.~S.~Manko,$^\dagger$ I.~M.~Mej\'ia,$^\ddagger$ C. I.
Ramos,$^\dagger$ and E.~Ruiz$^\sharp$}
\address{$^\dagger$Departamento de F\'\i sica, Centro de Investigaci\'on y
de Estudios Avanzados del IPN, A.P. 14-740, 07000 Ciudad de
M\'exico, Mexico\\$^\ddagger$Departamento de F\'\i sica, Escuela
Superior de F\'isica y Matem\'aticas, Instituto Polit\'ecnico
Nacional, 07738 Ciudad de M\'exico, Mexico\\$^\sharp$Instituto
Universitario de F\'{i}sica Fundamental y Matem\'aticas, Universidad
de Salamanca, 37008 Salamanca, Spain}

\begin{abstract}
In the present paper we study the singularity structure of the
exterior field of neutron stars with the aid of the four-parameter
exact solution of the Einstein-Maxwell equations. The complete
analysis of this problem in the generic case becomes possible due to
the implementation of the novel analytical approach to the
resolution of the singularity condition, and it shows the absence of
the ring singularities off the symmetry axis in the positive mass
case, as well as the possibility of the removal of the ring
singularity by a strong magnetic field in the negative mass case.
The solution takes an extraordinarily simple form in the equatorial
plane, very similar to that of the Kerr solution, which makes it
most suitable for astrophysical applications as the simplest model
of a rotating magnetized deformed mass. It also provides a
nontrivial example confirming a recent claim that the $\varphi$
component of the electromagnetic four-potential has features
inconsistent with the intrinsic properties of the electrovac metric,
while the magnetic field is represented correctly by the $t$
component of the dual electromagnetic four-potential.
\end{abstract}


\maketitle

\newpage

\section{Introduction}

The singularity structure of the Kerr solution \cite{Ker} describing
the exterior field of the ``most elementary'' rotating astrophysical
objects is well known and quite simple: the black-hole branch of the
solution possesses a ring singularity covered by the regular event
horizon, while in the hyperextreme sector of the solution,
characterized by the absence of the event horizon, the ring
singularity becomes ``naked'', i.e. visible to a distant observer.
The simplicity of one of the basic characteristics of the Kerr
spacetime is intrinsically and uniquely\footnote{The claim  about
nonuniqueness of the multipole moments of the Kerr solution recently
made in the paper \cite{BYa} is wrong.} defined by a specific set of
the relativistic Geroch-Hansen multipole moments \cite{Ger,Han,FHP},
concisely given by the formula $m_n=m(ia)^n$, $n=0,1,2,\dots$, the
parameter $m$ denoting the mass of the source and $a$ the angular
momentum per unit mass \cite{Han}. The above $m_n$ are coefficients
in the expansion of the function
\be
\xi(z)=\frac{1-e(z)}{1+e(z)}=\sum\limits_{n=0}^{\infty}\frac{m_n}{z^{n+1}},
\quad e(z)=\frac{z-m-ia}{z+m-ia}, \label{fax} \ee
when $z\to\infty$, and the knowledge of these $m_n$, calculated on
the symmetry axis, is sufficient to construct the corresponding
metric in the whole space by means of Sibgatullin's integral method
\cite{Sib,MSi}. The fact that all $m_n$ are functions of $m$ and $a$
only is to some extent reflected in the ``no-hair'' theorem
\cite{MTW} according to which the mass and angular momentum fully
characterize the Kerr black hole spacetime.

Next to black holes (BHs), neutron stars (NSs) are second densest
(and simplest) astrophysical objects in nature, and the exterior
field of NSs requires at least one more arbitrary physical
parameter, the mass quadrupole moment \cite{LPo}, which takes into
account the deformation of the source. An extensive study of the NS
models with the aid of the analytical and numerical approaches
carried out in recent decades
\cite{CST,SSu,SSu2,MMS,BSt,SFr,Ste,PAp,Pap2} has eventually led to a
remarkable discovery that the first few lowest multipole moments in
fact determine entirely the geometry around NSs, which constitutes
the essence of the Yagi {\it et al.} ``NS no-hair conjecture''
\cite{YKP}. The latter conjecture in turn naturally singles out the
six-parameter equatorially symmetric two-soliton solution \cite{MMR}
of the Einstein-Maxwell equations (henceforth referred to as MMR) as
a generic analytical model for the exterior geometry of a NS, which
also includes the parameters of electric charge and magnetic dipole
moment. For the pure vacuum specialization \cite{PAp2,MRu} of the
solution \cite{MMR} it has been shown \cite{MRu2} that the explicit
form of the multipole moments higher than quadrupoles can be read
off from the degeneration condition of the determinant $L_n$
involved in the description of the axis data defining the general
$N$-soliton vacuum metric \cite{MRu3}; the expressions for the
electromagnetic multipoles are obtainable from the expansions of the
electromagnetic potential via standard procedures
\cite{Sim,HPe,SAp,FCH,MMeR}.

One may expect that the singularity structure of the exact solutions
representing NSs, which are defined by the Yagi {\it et al.} NS
no-hair conjecture, is reacher compared to that of Kerr due to the
presence of some additional parameters and a more complicated form
of the respective metrical fields. Indeed, already the well-known
Tomimatsu-Sato $\delta=2$ (TS2) solution \cite{TSa}, which is a
particular case of the double-Kerr solution of Kramer and Neugebauer
\cite{KNe} and of the MMR spacetime and hence could in principle
describe the exterior of a specific NS, is endowed with a massless
ring singularity outside the symmetry axis accompanied by a region
with causality violation \cite{GRu}, and its origin was attributed
in the paper \cite{Man} to the presence of negative mass. Moreover,
the appearance of ring singularities in various particular 2-soliton
spacetimes has been routinely analyzed in a number of papers
\cite{MMS,MMSa,MRS,RMRS,CMR,MRS2}, where it has been shown, in
particular, that in the binary configurations of Kerr sources the
constituent with negative Komar \cite{Kom} mass develops a massless
ring singularity outside the symmetry axis which is needed to
prevent disintegration of that constituent, remembering that the
single Schwarzschild and Kerr sources of negative mass are known to
be unstable \cite{GHI,GDo}. Such ring singularities can also be
present when the two constituents have positive masses, in which
case they do not allow the dynamical non-regular evolution of the
joint stationary limit surfaces \cite{MRu4,MRu5}.

It may be observed that up to now the ring singularities arising in
binary systems and in the NS exterior have been usually analyzed
numerically because of the complexity of the explicit algebraic
expressions involved in the analysis. In this respect, it turns out
surprising that the singularity problem, as will be shown in the
present paper, can be solved analytically in the case of the
four-parameter subfamily of the MMR metric -- a NS solution
elaborated and studied a few years ago \cite{MRu5} that we are going
to rewrite in the extended parameter space in a manner slightly
different from the $e=0$ specialization of the electrovac metric
considered in \cite{MMeR}. Remarkably, this will permit us to
establish the absence of the massless ring singularities outside the
symmetry axis when the NS solution has positive mass, and also to
see that the ring singularity emerging in the negative mass case can
be removed by a strong magnetic field. Additionally, we shall obtain
a very simple form of the NS solution in the symmetry axis and
discuss the difference between the singularities of the metric
coefficients of the solution and those of the corresponding
component $A_\varphi$ of the electromagnetic four-potential.

Our paper is organized as follows. In the next section we will
consider the extended version of the solution \cite{MRu5} in which
the free parameters correspond to arbitrary relativistic multipole
moments, as well as the conditions defining the singularities of the
solution. In Sec.~III we solve analytically the condition for ring
singularities on and off the equatorial plane in the general case of
that solution, and compare the singularity structure of its extreme
vacuum limiting case with that of the well-known Tomimatsu-Sato
$\delta$=2 metric. The results obtained are discussed in Sec.~IV,
where in particular we present a very simple form of the NS solution
in the equatorial plane, and briefly comment on a better description
of the magnetic field by the $t$ component of the dual
electromagnetic potential than by the $\varphi$ component of the
usual electromagnetic four-potential.

\section{The extended 4-parameter solution for the NS exterior
and the singularity condition}

We remind that the NS solution \cite{MRu5} is determined by the axis
values of the Ernst complex potentials \cite{Ern} $\E$ and $\Phi$ of
the form
\bea \E(\rho=0,z)\equiv
e(z)=\frac{(z-m_1)(z-m_2)-ia(m_1+m_2)+a^2-\mu^2}
{(z+m_1)(z+m_2)+ia(m_1+m_2)+a^2-\mu^2}, \nonumber\\
\Phi(\rho=0,z)\equiv f(z)=\frac{i\mu(m_1+m_2)}
{(z+m_1)(z+m_2)+ia(m_1+m_2)+a^2-\mu^2}, \label{ado} \eea
where the parameters $m_1$, $m_2$, $a$ and $\mu$ take arbitrary real
values.

One can easily see that the mass parameters $m_1$ and $m_2$ in
(\ref{ado}) occur only in the combinations $m_1+m_2$ and $m_1m_2$,
which suggests that the extension of the parameter set can be
achieved by allowing for $m_1$ and $m_2$, in addition to the real
values they can take, to be also a pair of complex conjugate
quantities. In this way it turns out possible to introduce, instead
of $m_1$ and $m_2$, the total mass $m$ and the mass quadrupole
moment $Q$ of the source as arbitrary real parameters of the
solution. Indeed, since $m=m_1+m_2$ and $Q=-m(m_1m_2+a^2-\mu^2)$
\cite{MRu5}, then the substitution formulas take the form
\be m_1=(m+d)/2, \quad m_2=(m-d)/2, \quad
d=\sqrt{m^2+4(\k+a^2-\mu^2)}, \quad \k\equiv Q/m, \label{sub} \ee
where the quantity $d$ may take either real or pure imaginary
values, depending on the interrelations of the arbitrary parameters.
Then the axis data (\ref{ado}) assume the form
\be e(z)=\frac{z(z-m)-\k-ima}{z(z+m)-\k+ima}, \quad
f(z)=\frac{im\mu}{z(z+m)-\k+ima}, \label{adn} \ee
so that the four roots $\a_i$ of the algebraic equation
\be e(z)+\bar e(z)+2f(z)\bar f(z)=0, \label{SE} \ee
entering the final formulas of the solution, are given by
\be \a_1=-\a_2=\s_+, \quad \a_3=-\a_4=\s_-, \quad
\s_\pm=\sqrt{\textstyle{\frac{1}{4}}(m\pm d)^2-a^2+\mu^2}.
\label{spm} \ee
The corresponding Ernst potentials of the solution \cite{MRu5} hence
can be rewritten as
\bea \E&=&(A-B)/(A+B), \quad \Phi=C/(A+B), \nonumber\\
A&=&2\s_+\s_-[(m^2+d^2)(R_++R_-)(r_++r_-)-
2(m^2-d^2)(R_+R_-+r_+r_-)] \nonumber\\
&&-[(m+d)^2\s_-^2+(m-d)^2\s_+^2](R_+-R_-)(r_+-r_-) \nonumber\\
&&+4imad[\s_+(R_++R_-)(r_+-r_-)-\s_-(R_+-R_-)(r_++r_-)], \nonumber\\
B&=&4md\{\s_+\s_-[(m+d)(r_++r_-)-(m-d)(R_++R_-)] \nonumber\\
&&+ia[\s_+(m+d)(r_+-r_-)-\s_-(m-d)(R_+-R_-)]\}, \nonumber\\
C&=&4im\mu d[\s_-(m+d)(R_+-R_-)-\s_+(m-d)(r_+-r_-)], \nonumber\\
R_\pm&=&\sqrt{\rho^2+(z\pm\s_+)^2}, \quad r_\pm=\sqrt{\rho^2+(
z\pm\s_-)^2}, \label{EF1} \eea
and the extended metric functions from the line element
\be d s^2=f^{-1}[e^{2\gamma}(d\rho^2+d z^2)+\rho^2 d\varphi^2]-f(d
t-\omega d\varphi)^2 \label{Papa} \ee
have the form
\bea f&=&\frac{A\bar A-B\bar B+C\bar C}{(A+B)(\bar A+\bar B)}, \quad
e^{2\gamma}=\frac{A\bar A-B\bar B+C\bar C}{256d^4|\s_+|^2
|\s_-|^2R_+R_-r_+r_-}, \quad \omega=-\frac{{\rm Im}
[G(\bar A+\bar B)+C\bar I]}{A\bar A-B\bar B+C\bar C}, \nonumber\\
G&=&-2zB+2md\{\s_+[(m-d)^2+2\mu^2](R_++R_-)(r_+-r_-) \nonumber\\
&&-\s_-[(m+d)^2+2\mu^2](R_+-R_-)(r_++r_-)
-4imad(R_+-R_-)(r_+-r_-) \nonumber\\
&&-4\s_-[(m-d)\s_+^2-m\mu^2](R_+-R_-)
+4\s_+[(m+d)\s_-^2-m\mu^2](r_+-r_-) \nonumber\\
&&-4ia\s_+\s_-[(m-d)(R_++R_-)-(m+d)(r_++r_-)]\}, \nonumber\\
I&=&2im\mu\{[(m-d)\s_+^2+(m+d)\s_-^2](R_+-R_-)(r_+-r_-) \nonumber\\
&&-2\s_+\s_-[m(R_++R_-)(r_++r_-)-2(m+d)R_+R_--2(m-d)r_+r_-] \nonumber\\
&&-2iad[\s_+(R_++R_-)(r_+-r_-)-\s_-(R_+-R_-)(r_++r_-)] \nonumber\\
&&+2d\s_+\s_-[(3m+d)(R_++R_-)-(3m-d)(r_++r_-)+4md] \nonumber\\
&&+4imad[\s_-(R_+-R_-) -\s_+(r_+-r_-)]\}, \label{MF1} \eea
where $|x|^2$ means $x\bar x$. Note also that formulas for two
nonzero components of the electromagnetic four-potential remain the
same as in \cite{MRu5}:
\be A_t=-{\rm Re}\left(\frac{C}{A+B}\right), \quad A_\varphi={\rm
Im}\left(\frac{I-zC}{A+B}\right). \label{At} \ee

The Kerr solution is contained in the above formulas
(\ref{EF1})-(\ref{MF1}) as the particular case $\mu=0$, $\k=-a^2$.

Turning now to the analysis of the singularity structure of our
4-parameter solution outside the symmetry axis, we begin by
remarking that the ring singularities are roots of the algebraic
equation
\be A+B=0, \label{SPE} \ee
and these are located on the stationary limit surface (SLS) $f=0$ in
the pure vacuum case ($\mu=0$), or outside the SLS when the
electromagnetic field is present, similar to the case of the
Kerr-Newman solution \cite{New} endowed with negative mass (see,
e.g., Ref.~\cite{MRu6}). We find it plausible to first analyze the
appearance of ring singularities in the equatorial ($z=0$) plane,
their habitual location in the spacetimes with reflection symmetry
\cite{Kor,MNe,PSa,EMR}. This, on the one hand, will help the reader
understand a special character of the singularity structure of the
solution (\ref{EF1}) and, on the other hand, will simplify the
consideration of the general case where, as will be seen later on,
the singularities off the equatorial plane can only occur on the
symmetry axis.

As a preliminary, we first observe that in the equatorial plane
\be R_+=R_-=\sqrt{\rho^2+\s_+^2}, \quad
r_+=r_-=\sqrt{\rho^2+\s_-^2}, \label{rz0} \ee
and the condition (\ref{SPE}) takes the form
\be [(m-d)R_--(m+d)r_-][(m+d)R_--(m-d)r_-+2md]\s_+\s_-=0. \label{sc}
\ee
The above condition will be fulfilled if one of the factors on the
left-hand side of (\ref{sc}) takes zero value. In what follows we
exclude the cases $\s_+=0$ and $\s_-=0$ as leading to the double
roots of Eq.~(\ref{SE}) and hence to indetermination of the
potentials $\E$ and $\Phi$ in (\ref{EF1}). The remaining two
conditions to investigate are therefore
\be (m-d)R_--(m+d)r_-=0, \label{cond1} \ee
and
\be (m+d)R_--(m-d)r_-+2md=0. \label{cond2} \ee
Mention that out of these two it is the condition (\ref{cond1}) that
gives rise to the ring singularity outside the symmetry axis
independently of the sign of the mass parameter $m$, the location of
the singularity being defined by
\be \rho_S=\sqrt{a^2-\mu^2}, \quad z_S=0, \quad a^2>\mu^2.
\label{rzs} \ee
In particular, the above formula is verified by the singularity
shown in Fig.~3 of \cite{MRu5} for the parameter choice $m=2.5$,
$a=0.5$, $\mu=0.25$.

Remarkably, after a considerable effort spent by us on identifying
in a rigorous way of all the cases when the singularity (\ref{rzs})
is present in the solution, we have eventually discovered, to our
big surprise, that the condition (\ref{cond1}) is not in fact
involved in the analysis of the singularity problem because the
complex potentials (\ref{EF1}) degenerate in the equatorial plane to
the expressions
\be \E=\frac{(m+d)R_--(m-d)r_--2md}{(m+d)R_--(m-d)r_-+2md}, \quad
\Phi=0, \label{EF1e} \ee
due to appearance, at $z=0$, of the factor $(m-d)R_--(m+d)r_-$ both
in the denominator and numerator of $\E$. As a result, the
singularity problem in the equatorial plane considerably simplifies
and reduces to analyzing the condition (\ref{cond2}) alone.

\section{Solving the singularity problem in and off the equatorial plane}

Fortunately, the analysis of the condition (\ref{cond2}) is
straightforward. Indeed, solving (\ref{cond2}) for $r_-$, we get
\be r_-=\frac{R_-(m+d)+2md}{m-d}, \label{rm} \ee
and this must be substituted into the equality
\be R_-^2-r_-^2-\s_+^2+\s_-^2=0, \label{eqRr} \ee
which is a direct consequence of (\ref{rz0}), thus leading to the
equation
\be -\frac{md(2R_-+m+d)^2}{(m-d)^2}=0, \label{cRm} \ee
with the obvious solution
\be R_-=-\frac{1}{2}(m+d). \label{Rmf} \ee
Then the substitution of (\ref{Rmf}) into (\ref{rm}) gives
\be r_-=-\frac{1}{2}(m-d), \label{rmf} \ee
so that (\ref{Rmf}) and (\ref{rmf}) define the location of the
singularity in the equatorial plane.

Our further estimations must take into account that all four square
roots $R_\pm$ and $r_\pm$ entering the solution (\ref{EF1}) are
understood as the positive branch of these functions: ${\rm
Re}(R_\pm)>0$, ${\rm Re}(r_\pm)>0$. We should also bear in mind that
the quantity $d$ can assume (positive) real or pure imaginary
values, i.e. distinguish between the cases $d^2>0$ and $d^2<0$. Then
the former case implies that the inequalities
\be m+d<0 \quad {\rm and} \quad m-d<0 \label{in1} \ee
must be satisfied simultaneously, whence we get immediately
\be m<0. \label{ml0} \ee
On the other hand, the latter case $d^2<0$ also leads directly to
the condition (\ref{ml0}) by applying the above mentioned criterion
on the square roots to the expressions (\ref{Rmf}) and (\ref{rmf}).
Therefore, we can conclude that the positive values of $m$ in our
solution do not develop singularities in the equatorial plane.

The value of $\rho$ in the $z=0$ plane at which the singularity
occurs in the negative mass case is readily obtainable from
(\ref{rz0}), (\ref{Rmf}) and (\ref{rmf}), and it is given by the
above formula (\ref{rzs}) for $\rho_S$. It suggests that negative
mass of the source itself is not yet a guarantee for the formation
of a ring singularity. Indeed, it is clear that for the values
$\mu^2>a^2$ of the magnetic dipole parameter $\mu$ the singularity
foes not arise, which means that the magnetic field plays a
stabilizing role in the NS solution.

Another restriction on the singularity in the $m<0$ case follows
from the first inequality in (\ref{in1}):
\be d<-m \quad \Longrightarrow \quad \k<\mu^2-a^2, \label{in2} \ee
which means in particular that the oblate configurations of the
negative mass source in our solution, corresponding to positive
$\k$, do not develop a ring singularity in the equatorial plane.
Recalling that the Kerr solution with negative mass is always
accompanied by a massless ring singularity outside the symmetry axis
\cite{Man}, we can draw a conclusion that NSs might probably be more
stable negative mass configurations than the $m<0$ Kerr source due
to presence of an arbitrary mass quadrupole moment.

\subsection{The extreme case}

We find it very instructive to compare the singularity structure of
the extreme case of the solution (\ref{EF1}) with that of the
Tomimatsu-Sato $\delta=2$ spacetime \cite{TSa} which was
historically the first nontrivial example of an asymptotically flat
stationary axisymmetric solution constructed within the framework of
the Ernst formalism \cite{Ern2}. Written in the prolate ellipsoidal
coordinates ($x,y$), the Ernst potential of the TS2 solution has the
form
\bea &&\E=(A-B)/(A+B), \nonumber\\
&&A=p^2x^4+q^2y^4-1-2ipqxy(x^2-y^2), \nonumber\\
&&B=2px(x^2-1)-2iqy(1-y^2), \nonumber\\ &&x=\frac{1}{2k}(r_++r_-),
\quad y=\frac{1}{2k}(r_+-r_-), \quad r_\pm=\sqrt{\rho^2+(z\pm k)^2},
\label{ETS} \eea
where the real constants $p$ and $q$ are subject to the constraint
$p^2+q^2=1$, and $k$ is an arbitrary real positive parameter. The
total mass of this solution is $M=2k/p$, so, with account of the
positivity of $k$, the positive or negative values of $M$ are
determined by the positive or negative values of $p$, respectively.
As had been pointed out by Tomimatsu and Sato themselves, their
solution (for $q\ne0$) always has a ring singularity in the
equatorial plane ($y=0$), the exact location of which in the
positive-mass case being given by the formula \cite{Man}
\be
x_0=\frac{1}{2p}\left(\chi-1+\sqrt{3-\chi^2-\frac{2}{\chi}(1-2p^2)}
\right), \quad y_0=0, \quad
\chi\equiv\sqrt{1+(2p)^{2/3}(p^2-1)^{1/3}}, \label{sTS} \ee
while for the locus of the singularity in the negative-mass case we
refer the reader to Ref.~\cite{Man}.

On the other hand, the special case $\s_+=\s_-$, or $d=0$,
corresponding to a pair of equal $\a$'s in (\ref{spm}) was worked
out in \cite{MRu5}, and in the absence of the magnetic field
($\mu=0$) it is defined by the Ernst potential of the form
\bea &&\E=(A-B)/(A+B), \nonumber\\
&&A=m^4(x^4-1)+a^4(x^2-y^2)^2-2m^2ax[ax(x^2-y^2)-2i\s y(1-y^2)],
\nonumber\\
&&B=2m\{m^2[\s x(x^2-1)+iay(1-y^2)]-a\s(x^2-y^2)(ax-i\s y)\},
\nonumber\\ &&x=\frac{1}{2\s}(r_++r_-), \quad
y=\frac{1}{2\s}(r_+-r_-), \quad r_\pm=\sqrt{\rho^2 +(z\pm \s)^2},
\quad \s=\sqrt{m^2-a^2}, \label{EMR} \eea
where the parameter $m$ is now half the $m$ that appears in the
general solution (\ref{EF1}). It was shown in \cite{MeMR} that the
subextreme branch of the solution (\ref{EMR}) belongs to the
Kinnersley-Chitre family of the vacuum spacetimes \cite{KCh}.

It is easy to check that, similar to the generic case, the Ernst
potential (\ref{EMR}) considerably simplifies in the equatorial
plane, taking the form
\be \E=\left(\frac{\s x-m}{\s x+m}\right)^2, \label{MRep} \ee
and hence the singularity occurs at
\be x=-m/\s, \label{MRsin} \ee
so that the ring singularity outside the symmetry axis requiring
$x>1$ (i.e. $\rho>0$) can only be developed by a negative mass. It
is worth noting that in the case of the TS2 solution a similar
simplification of the Ernst potential does not take place, and we
have
\be \E_{TS}(y=0)=\frac{p^2x^4-2px(x^2-1)-1}{p^2x^4+2px(x^2-1)-1},
\label{TSep} \ee
that is the quartic polynomials in $x$ both in the numerator and
denominator of $\E$, like in the general formulas (\ref{ETS}).

Fortunately, the singularity problem of the extreme solution
(\ref{EMR}) can be solved in the entire space too due to exceptional
factorization  properties of this solution. In the generic case, the
system of equations to be solved is
\be {\rm Re}(A+B)=0, \quad {\rm Im}(A+B)=0, \label{con1} \ee
and the first equation, as it follows from (\ref{EMR}), takes the
form
\be (\s^2 x^2+a^2y^2-m^2)[(\s x+m)^2+a^2y^2]=0, \label{con1} \ee
while the second equation yields
\be 2amy[(\s x+m)(\s x-2my^2+m)+a^2y^2]=0. \label{con2} \ee
Actually, we must be only interested in the first factor in
(\ref{con1}) and the last factor in (\ref{con2}) since the condition
$y=0$ leads to the ``equatorial'' case already considered above,
whereas the second factor on  the left-hand side of (\ref{con1}) is
a positive quantity for $y\ne0$. Moreover, if in addition, instead
of equating to zero the last factor in (\ref{con2}), we shall
consider the difference of that factor with the first factor in
(\ref{con1}), then the system to be solved becomes eventually
composed of the following two equations:
\be \s^2x^2+a^2y^2-m^2=0, \quad (1-y^2)(\s x+m)=0, \label{sysf1} \ee
and besides the solution $y=0$, $x=-m/\s$ involving the equatorial
plane, we arrive at the off-plane solutions
\be y=1, \quad x=\pm1 \quad \mbox{and} \quad y=-1, \quad x=\pm1.
\label{solOP} \ee
Apparently, these solutions do not represent the ring singularities
outside the symmetry axis.

\subsection{The general case}

It is remarkable that the generic analysis of the ring singularities
outside the symmetry axis admits in the case of the solution
(\ref{EF1}) a purely analytical treatment. However, since it does
not lead to the new singularities in addition to the already
considered in the equatorial plane, thus only confirming their
absence off the latter plane, in what follows we shall restrict
ourselves to the description of the general scheme of our novel
approach to the resolution of the condition (\ref{SPE}), omitting
numerous particular subcases which, even being of some utility
mathematically, do not provide us with an extra physical
information. As a sort of a compensation, we shall also consider a
curious case of a line singularity on the symmetry axis covered by
our general scheme for a special relation of the parameters of the
solution.

Obviously, the system of algebraic equations defining the
singularities on and off the equatorial plane that must be solved by
us is composed of the conditions for vanishing real and imaginary
parts of (\ref{SPE}), i.e.,
\bea &&2\s_+\s_-\{[(m^2+d^2)(R_++R_-)(r_++r_-)-
2(m^2-d^2)(R_+R_-+r_+r_-)] \nonumber\\
&&\hspace{0.8cm}+2md[(m+d)(r_++r_-)-(m-d)(R_++R_-)]\} \nonumber\\
&&\hspace{0.8cm}-[(m+d)^2\s_-^2+(m-d)^2\s_+^2](R_+-R_-)(r_+-r_-)=0,
\label{eq1} \eea
and
\be \s_+(r_+-r_-)(R_++R_-+m+d)-\s_-(R_+-R_-)(r_++r_-+m-d)=0.
\label{eq2} \ee

Note that one also can easily arrive at the system
(\ref{eq1})-(\ref{eq2}) by merely exploring the equatorial symmetry
of the solution (\ref{EF1}), which implies that if the point
$(\rho_S,z_S)$ satisfies Eq.~(\ref{SPE}) then the point
$(\rho_S,-z_S)$ will satisfy this equation too.

Though the resolution of the above system may look impossible
analytically since the dependence in it on the coordinates
$(\rho,z)$ is not direct but through the square roots $R_\pm$ and
$r_\pm$, we still have been able to find a way out of this
unpleasant situation by considering that $R_\pm$ and $r_\pm$ enter
Eqs.~(\ref{eq1}) and (\ref{eq2}) as independent variables, which
requires to supplement the latter equations with two additional
conditions/constraints that must verify $R_\pm$ and $r_\pm$ as
functions of $\rho$ and $z$. Thus, using the definition (\ref{EF1})
of $R_\pm$ and $r_\pm$, we can express $\rho$ and $z$, say, in terms
of $R_\pm$ as
\be \rho^2=R_-^2-\frac{(R_-^2-R_+^2+4\s_+^2)^2}{16\s_+^2}, \quad
z=\frac{R_+^2-R_-^2}{4\s_+}, \label{rz} \ee
and then, with the aid of $r_\pm$, we get the following two extra
conditions
\bea (\s_+-\s_-)R_+^2+(\s_++\s_-)R_-^2-2\s_+(r_-^2+\s_+^2-\s_-^2)=0,
\nonumber\\
(\s_+-\s_-)R_-^2+(\s_++\s_-)R_+^2-2\s_+(r_+^2+\s_+^2-\s_-^2)=0,
\label{ac} \eea
which complement the system (\ref{eq1})-(\ref{eq2}). Of course, once
the particular values of the functions $R_\pm$ and $r_\pm$ defining
the singularities are found, the corresponding $(\rho_S,z_S)$ should
be obtained by means of formulas (\ref{rz}).

The advantage of solving the system of four equations (\ref{eq1}),
(\ref{eq2}), (\ref{ac}) for $R_\pm$ and $r_\pm$ instead of directly
solving the two equations (\ref{eq1}) and (\ref{eq2}) for $(\rho,z)$
turns out significant, as now we can obtain from (\ref{eq1}) and
(\ref{eq2}), via standard substitutions, the following four
analytical solutions for $R_\pm$, $r_\pm$:

{\it Solution 1.}
\bea R_\pm&=&\frac{1}{2}[h(m-d)(\pm2k\s_+-1)-m-d], \nonumber\\
r_\pm&=&\frac{1}{2}[h(m+d)(\pm2k\s_--1)-m+d]. \label{sol1} \eea

{\it Solution 2.}
\bea R_\pm&=&-\frac{1}{2}(m+d)[h(\pm2k\s_+-1)(4k^2\s_-^2-1)+1], \nonumber\\
r_\pm&=&-\frac{1}{2}(m-d)[h(\pm2k\s_--1)(4k^2\s_+^2-1)+1].
\label{sol2} \eea

{\it Solution 3.}
\bea R_\pm&=&\frac{1}{2}[\pm2h(m-d)\s_+-m-d], \nonumber\\
r_\pm&=&\frac{1}{2}[\pm2h(m+d)\s_--m+d]. \label{sol3} \eea

{\it Solution 4.}
\bea R_\pm&=&-\frac{1}{2}(m+d)(\pm2h\s_-+1), \nonumber\\
r_\pm&=&-\frac{1}{2}(m-d)(\pm2h\s_++1). \label{sol4} \eea

The above solutions contain the arbitrary (real or complex)
constants $h$ and $k$, which in our scheme would be better regarded
as arbitrary variables, whose particular form should be found from
the constraint equations (\ref{ac}). The evaluation of these $h$ and
$k$ in each case is straightforward and does not represent any
difficulty. At the same time, a thorough analysis of the solutions
(\ref{sol1})-(\ref{sol4}) can be shown to lead to the  ring
singularities located in the equatorial plane only, and hence we can
skip its presentation here since the description of such
singularities has already been given in the first part of this
section. We only would like to remark in this respect that some of
the ``equatorial'' solutions resulting from the general scheme must
be discarded after accounting for the degeneration of the potentials
(\ref{EF1}) in the equatorial plane.

A physically interesting nontrivial case still worthy of mentioning
in relation with our general scheme arises after the substitution of
the solution (\ref{sol3}) into the constraint equations and noting
that these are satisfied when $\mu^2=a^2$, independently of the
value of $h$. In this particular case the expressions of $\s_\pm$
simplify to the form
\be \s_\pm=\frac{1}{2}\epsilon_\pm(m\pm d), \quad \epsilon_\pm^2=1,
\label{spm5} \ee
so that the corresponding $R_\pm$ and $r_\pm$ become
\bea R_\pm&=&\frac{1}{2}(m+d)[\pm\epsilon_+h(m-d)-1], \nonumber\\
r_\pm&=&\frac{1}{2}(m-d)[\pm\epsilon_-h(m+d)-1], \label{Rr51} \eea
or, equivalently, after expressing $m$ and $d$ in terms of $\s_\pm$
with the aid of (\ref{spm5}),
\bea R_\pm&=&\s_+(\pm 2h\epsilon_-\s_--\epsilon_+), \nonumber\\
r_\pm&=&\s_-(\pm 2h\epsilon_+\s_+-\epsilon_-). \label{Rr52} \eea

Now we can identify the location of the singularity by substituting
$R_\pm$ from (\ref{Rr52}) into (\ref{rz}), yielding
\be \rho=0, \quad z=-2h\epsilon_+\epsilon_-\s_+\s_-. \label{rzsin}
\ee

Using the explicit form (\ref{spm5}) for $\s_\pm$ in the $\mu^2=a^2$
case, (\ref{rzsin}) also rewrites as

\be \rho=0, \quad z=2h\kappa, \label{rzs2} \ee
which shows in particular that the singularity's locus reduces to
the origin $\rho=z=0$ in the case of vanishing $\kappa$. However,
when $\kappa\ne0$, one could think that the singularity extends
along the whole symmetry axis, as $h$ can take arbitrary real
values, thus suggesting that the solution (\ref{EF1}) might be not
asymptotically flat at least in the special $\mu^2=a^2$ case.

The issue of the above singularity can be clarified after recalling
that the roots $R_\pm$ and $r_\pm$ are defined under the positive
branch criterion. Therefore, the values (\ref{Rr52}) which, after
changing $h$ to $z$ by means of (\ref{rzsin}), take the form
\be R_\pm=\mp\epsilon_+(z\pm\s_+), \quad
r_\pm=\mp\epsilon_-(z\pm\s_-), \label{Rr5a} \ee
must be consistent with the axis values $R_\pm(0,z)$, $r_\pm(0,z)$
that follow from their definition (\ref{EF1}), i.e.,
\be R_\pm(0,z)=\sqrt{(z\pm\s_+)^2}, \quad
r_\pm(0,z)=\sqrt{(z\pm\s_-)^2}. \label{Rra} \ee

Then, for instance, if $\s_\pm$ are real-valued and $\s_+>\s_->0$,
we get on the upper part of the symmetry axis ($z>\s_+$)
\be R_\pm(0,z)=z\pm\s_+, \quad r_\pm(0,z)=z\pm\s_-, \label{Rrau} \ee
and a simple inspection shows that there is no choice of
$\epsilon_\pm$ for which $R_\pm$ and $r_\pm$ in (\ref{Rr5a}) would
fully coincide with $R_\pm(0,z)$ and $r_\pm(0,z)$ in (\ref{Rrau}).
The same is true for the lower part of the symmetry axis
($z<-\s_+$), and also for the parts $\s_-<z<\s_+$ and
$-\s_+<z<-\s_-$. However, on the intermediate part $-\s_-<z<\s_-$ of
the $z$-axis, formulas (\ref{Rra}) assume the form
\be R_\pm(0,z)=\pm z+\s_+, \quad r_\pm(0,z)=\pm z+\s_-, \label{Rri}
\ee
and now the choice $\epsilon_+=\epsilon_-=-1$ makes the expressions
(\ref{Rr5a}) and (\ref{Rri}) identical, which means that the
singularity defined by formulas (\ref{rzs2}) occupies exclusively
the finite interval ($-\s_-,\s_-$) of the symmetry axis. In
addition, it is not difficult to see that this singularity is
developed by the {\it negative} mass. Indeed, under the above choice
$\epsilon_\pm=-1$ and our supposition $\s_\pm>0$, formulas
(\ref{spm5}) imply that
\be m+d<0, \quad m-d<0, \label{ineq5} \ee
whence we immediately arrive at $m<0$.

\section{Discussion}

The analysis carried out in the present paper shows that the
singularity structure of the exterior field of NSs defined by the
solution (\ref{EF1}) is quite similar to that of black holes: no
ring singularities are present outside the symmetry axis in the
positive mass case, and a ring singularity located in the equatorial
plane arises in the case of negative mass. To some extent, the
singularity of NSs in the latter case has a more benign character
than that of black holes because in the Kerr and Kerr-Newman
solutions with negative mass the singularity is irremovable
\cite{Man,MRu6}, while a strong magnetic field $\mu^2>a^2$ in the
solution (\ref{EF1}) removes the ring singularity. This could be
interpreted, bearing in mind the generic instability of the negative
mass sources \cite{GHI,GDo} (the singularities thus preserving the
stationarity of the sources), as a stabilization effect exerted by
the magnetic field on the massive sources of NSs. In other words,
NSs carrying negative mass and magnetic dipole moment are more
stable objects than the ``black holes'' of negative mass.
Interestingly, the position of the ring singularity defined by
formula (\ref{rzs}) does not depend on the mass quadrupole parameter
$\kappa$.

The analytical approach to the singularity problem of NSs developed
in the present paper has allowed us to rectify some incorrect
statements about the presence of the ring singularities made earlier
in the literature \cite{MRu5} on the basis of the numerical analysis
of the condition (\ref{SPE}). As a matter of fact, the idea to
resort to analytical study of the singularity problem came to us
only after we were able to clearly realize that the numerical
methods were failing in the vicinity of singular points and were
often producing some exotic unrealistic results that could be
erroneously taken for the genuine ring singularities.

An important outcome of our consideration is a surprisingly simple
form of the neutron star metric (\ref{MF1}) in the equatorial plane
($z=0$). Thus, for the metric functions $f$ and $\omega$ taking part
in the analysis of the behavior of test particles and study of
various phenomena occurring in this plane, formulas (\ref{MF1}) give
us the following simple expressions:
\bea f&=&\frac{A-B}{A+B}, \quad \omega=-\frac{W}{A+B}, \nonumber\\
A&=&(m+d)r_+-(m-d)r_-, \quad B=2md, \quad W=4mad, \nonumber\\
r_\pm&=&\sqrt{\rho^2+\textstyle{\frac{1}{2}}(m^2+2\kappa\pm md)},
\quad d=\sqrt{m^2+4(\kappa+a^2-\mu^2)}, \label{fw} \eea
which are by far simpler than the analogous expressions for $f$ and
$\omega$ obtained in the papers \cite{SSu,MRu} (pure vacuum case)
and \cite{Man2} (the electrovac case); (\ref{fw}) also improve and
generalize the ``equatorial'' formulas of the paper \cite{MeMR} for
which the possibility of further simplifications was previously
overlooked. Actually, the above formulas (\ref{fw}) are practically
as simple as in the case of the Kerr solution due to the linear
dependence of $A$ on $r_\pm$ and constant values of $B$ and $W$, the
latter black hole case being contained in (\ref{fw}) just as the
$d=m$ (i.e., $\kappa=-a^2$, $\mu=0$) specialization. At the same
time, compared to the Kerr case, Eqs.~(\ref{fw}) represent much more
generic sources, as they contain two additional independent
parameters $\kappa$ and $\mu$ that allow one to take account of an
arbitrary quadrupole mass deformation and of the dipole magnetic
field of neutron stars. As a curiosity, we find it worth noting that
the constant object $d$ entering (\ref{fw}) can take arbitrary real
or pure imaginary values, leaving the functions $f$ and $\omega$ to
be real-valued in both cases.

It is also remarkable that the solution (\ref{EF1}) can be
considered as providing a nontrivial evidence in favor of the recent
claim \cite{RGM,GMR} that the component $A_\varphi$ of the
electromagnetic four-potential does not describe correctly the
magnetic field of the Einstein-Maxwell spacetimes. Although in the
aforementioned papers the claim was made about the asymptotically
nonflat potential $A_\varphi$ linked to the magnetic charge and
possessing two semi-infinite singularities, so that the claim looks
quite natural, the function $A_\varphi$ of the solution (\ref{EF1}),
on the other hand, is defined by formula (\ref{At}) and is
asymptotically flat, thus presumably looking well-behaved and
reflecting correctly the properties of the metric (\ref{MF1}).
However, our study of the singularities of the solution (\ref{EF1})
in the equatorial plane makes it possible to establish a not very
conspicuous inconsistency between the singularity structures of the
metric (\ref{MF1}) and the corresponding potential $A_\varphi$.
Indeed, it follows from the formulas (\ref{EF1}), (\ref{MF1}) and
(\ref{At}) that in the equatorial plane the potential $A_\varphi$
takes the form
\be A_\varphi=\frac{2m\mu(r_+-r_-+d)}{(d-m)r_++(d+m)r_-},
\label{Afe} \ee
where $r_\pm$ and $d$ are the same as in (\ref{fw}). The above
formula (\ref{Afe}) means that the simplification of the expression
of $A_\varphi$ at $z=0$ occurs differently than in the case of the
functions $f$ and $\omega$: while the formulas (\ref{fw}) have been
obtained after canceling out the common factor $(d-m)r_++(d+m)r_-$
in the numerators and denominators of $f$ and $\omega$, the common
factor leading to (\ref{Afe}) is $(m+d)r_+-(m-d)r_-+2md$. As a
result, the singularity structure of $A_\varphi$ in (\ref{Afe}) is
defined by the roots of Eq.~(\ref{cond1}), whereas the singularities
of $f$ and $\omega$ in (\ref{fw}) emerge as the roots of
Eq.~(\ref{cond2}). Therefore, the singularity structure of
$A_\varphi$ turns out to be inconsistent with that of the metric
functions $f$ and $\omega$ because, as we have already mentioned,
the singularity (\ref{rzs}) satisfies the condition (\ref{cond1})
for the positive values of $m$ too. The correct description of the
magnetic field in the solution (\ref{EF1}) is provided by the
well-behaved $t$ component $B_t={\rm Im}(\Phi)$ of the dual
electromagnetic four-potential $B_\mu$ which is devoid of the
undesirable singularities of the potential $A_\varphi$ since,
according to (\ref{EF1e}), $B_t$ vanishes in the equatorial plane.

As a final remark we would like to mention that the similarity in
the singularity structures of the black-hole and NS solutions looks
also extending to their multipole structures. Indeed, the complex
moments $m_n$ determining the mass and angular momentum
distributions of the NS solution obtainable from (\ref{fax}) and
(\ref{adn}) have a very simple form
\be m_{2n}=m\kappa^n, \quad m_{2n+1}=ima\kappa^n, \quad n=0,1,2,...
\label{mmg} \ee
quite comparable in simplicity with those of the Kerr solution,
while for the electromagnetic moments $q_n$ arising as coefficients
in the expansion ($z\to\infty$)
\be
\eta(z)=\frac{2f(z)}{1+e(z)}=\sum\limits_{n=0}^{\infty}\frac{q_n}{z^{n+1}},
\label{eta} \ee
we get from (\ref{adn}) and (\ref{eta}) the expressions
\be q_{2n}=0, \quad
q_{2n+1}=im\mu\kappa^n, \quad n=0,1,2,... \label{mme} \ee
which even exceed in simplicity the corresponding $q_n$ of the
Kerr-Newman black hole whose $q_{2n}$ are all nonzero.

\section*{Acknowledgments}

One of us (VSM) would like to thank Erasmo G\'omez for technical
computer assistance. This work was partially supported by CONACyT of
Mexico, by Project SA096P20 from Junta de Castilla y Le\'on of
Spain, and by Project PGC2018-096038-B-100 funded by the Spanish
``Ministerio de Ciencia e Innovaci\'on'' and FEDER ``A way of making
Europe''.

\end{document}